\documentstyle[amssymb,psfig,aps]{revtex}

\draft

\begin{document}

\tighten

\twocolumn[\hsize\textwidth\columnwidth\hsize\csname@twocolumnfalse\endcsname

\title{Ground state properties of a trapped few-Boson system under rotation\\
---beyond the ``lowest Landau level'' approximation---}
\author{Xiaji Liu$^{1,3}$, Hui Hu$^1$, Lee Chang$^2$, and Shi-Qun Li$^{1,3}$}
\address{$^1$Department of Physics, Tsinghua University, Beijing 100084, China\\
$^2$Center for Advanced Study, Tsinghua University, Beijing 100084, China\\
$^3$Key Laboratory for Quantum Information and Measurements Ministry of Education\\
and Center for Atomic and Molecular Nanosciences, Tsinghua University, Beijing 100084, China}
\date{\today}
\maketitle

\begin{abstract}
We consider a harmonically trapped few-Boson system under rotation and investigate the 
ground state properties beyond the usual ``lowest Landau level'' approximation by using
exact diagonalizations in a restricted Hilbert subspace. We find that both the effective
interaction energy and density distribution are strongly affected by the two-body 
interaction strength.
\end{abstract}


]

Following the experimental realization of Bose-Einstein condensation of
alkali-metal atoms \cite{review}, there has been much attention attached to
the behavior of these systems under rotation both experimentally and
theoretically. Mattews {\it et al.} \cite{matthews} created vortices in a
two-component system, and Madison {\it et al.} \cite{madison} studied the
rotations in a stirred one-component Bose-Einstein condensate. Theoretical
studies have mainly focused on the Thomas-Fermi regime of strong
interactions \cite{rokhsar,sinha,feder,fetter,marzlin} or on the limit of
weak interactions between the atoms \cite
{wilkin98,mft,mottelson,bertsch,wilkin00,mft-vs-ex,analytic}. In the latter
case, most research works have applied the ``lowest Landau Level (LLL)''
approximation, in which all the bosons are in single-particle orbitals with
radial quantum number $n=0$ and angular momentum $m$ having the same sign as
the total angular momentum. The intermediate regime of moderate interaction,
however, is rarely concerned \cite{ahsan,isoshima}.

In this Brief Report, we would like to study ground state properties of a
trapped rotating Bose-Einstein condensate with an {\em arbitrary} two-body
interaction strength beyond the LLL approximation. To attack the problem, we
calculate low-lying states of a few-Boson system by using the exact
diagonalization method. We assume that our numerical computation is {\em %
qualitatively }applicable for a rotating Bose-Einstein condensate though the
number of bosons is limited to a small value of $N=16$. We also assume that
the cloud of atoms rotates about some axis, and that the system is in its
ground state with respect to this axis, which implies that our problem is
essentially two dimensional.

The trapped Bose-Einstein condensate, comprised of $N$ repulsively
interacting alkali atoms with mass $M$ and s-wave scattering length $a_{sc}$%
, obeys the many-body Hamiltonian 
\begin{equation}
{\cal H}=\sum\limits_{j=1}^N\left\{ -\frac 12{\bf \nabla }_j^2+\frac 12{\bf r%
}_j^2\right\} +g\sum\limits_{i<j}2\pi \delta ({\bf r}_i-{\bf r}_j),
\end{equation}
where the energy and length are given throughout in scaled harmonic
oscillator units $\hbar \omega $ and $a=\sqrt{\hbar /M\omega }$,
respectively. The two-body contact interaction strength is characterized by
a dimensionless coupling constant $g=a_{sc}/a$. As we shall see, the problem
has three different interesting regimes parametrized by the product $gN$:
the Thomas-Fermi regime $gN\gg 1$, the moderate interaction regime $gN\sim 1$%
, and the weak interaction regime $gN\ll 1$. To diagonalize the Hamiltonian,
Eq. (1), the single-particle states of the Fock state are chosen to be
eigenstates of the two-dimensional harmonic oscillator, i.e., 
\begin{eqnarray}
\epsilon _{nm} &=&2n+\left| m\right| +1, \\
\psi _{nm} &=&\sqrt{\frac{n!}{\pi (n+\left| m\right| )!}}r^{\left| m\right|
}e^{-\frac{r^2}2}L_n^{\left| m\right| }(r^2)e^{im\varphi },
\end{eqnarray}
where $L_n^{\left| m\right| }(r^2)$ is the associated Laguerre polynomial. $%
n $ and $m$ are the radial and angular momentum quantum number,
respectively. Previous studies in weak interaction limit usually take single
particle states with $n=0$ and $m=0,+1,+2,+3,$..., which corresponds to the
LLL approximation. Here we set up Fock states by extending to $n\leqslant 2$
and $-2\leqslant m\leqslant 4$ \cite{nm-note}, and sample over the full
Hilbert space with a fixed number of bosons. From this sampling, only those
Fock states with a given total orbital angular momentum $L$ and a
configuration energy (corresponding to the sum of occupied single particle
energies) less than or equal to a specified cutoff energy $E_c$ are included
(see figure 1 below for example). The purpose was to select only the most
important Fock states from the full basis, thereby reducing the matrix
dimension to an acceptable size $d\lesssim 10^5$. Once the active Fock
states are constructed, we calculate the matrix elements and subsequently
diagonalize the matrix by using the Davidson algorithm \cite{davidson},
which is very efficient to solve the eigenvalue problem with a large and
sparse matrix.

It is important to point out that the above diagonalization scheme only in
principle yields an exact solution of the many-body problem. For reasons of
numerical feasibility it is necessary to truncate the set of basis functions
to be used in the diagonalization. One then has to make sure that
convergence of the ground state energy is reached with respect to the
cutoff. As the required matrix size increases rapidly with $gN$ and $L$,
computational expenses severely restrict the calculations to only the
smallest systems at not too large values of $g$. Thus, with increasing boson
number or $g$, the results become less accurate due to the restricted number
of basis states that can be included in the calculations. Figure 1 shows the
convergence of the ground state energy as a function of the cutoff energy $%
E_c$ for a system of $N=10$ bosons at $L=10$. The lowest possible Fock state
has all ten bosons distributed in the lowest Landau level, and the
corresponding configuration energy equals to $N+L=20$. This means that for
the ground state energy with different cutoff energies displayed in figure 1
all excitations up to an energy $E_c-20$ are included. It is readily seen
that both the case with a moderate interaction strength $gN=1$ (figure 1a)
and a relatively strong interaction strength $gN=3$ (figure 1b) show
convergence at $E_c>30$. The smaller $g$ is, the more rapidly the convergene
reaches. Therefore, we conclude that our numerical results shown below is
very accurate for a weak or moderate interaction strength, and also
qualitatively accurate in the relatively strong interaction regime. In what
follows, we take $E_c=N+L+10.$

We now analyze the influence of interaction strength $g$ on ground state
energies as a function of the total angular momentum, which may be written
as 
\begin{equation}
E_{gs}(L)=(N+L)+gV_{int,L},
\end{equation}
where $V_{int,L}$ is introduced as a {\em scaled} effective interaction
energy. Figure 2 shows the $L$ dependence of $V_{int,L}$ of a system of $%
N=16 $ bosons for $g=0.006,$ $0.1$ and $0.3$. For comparison, the result
with LLL approximation is also displayed. Note that the case of $g=0.006$ is
closer to experimental values for $^{87}$Rb \cite{rb}, i.e., $a_{sc}\sim
100a_0=5.29 $ nm and $a=1.25\times 10^{-4}$ cm. For all three interaction
strengths, the interaction energy $V_{int,L}$ simply decreases linearly or
near linearly with increasing the angular momentum. At $N=L$, a kink appears
in slope. This is a hint of condensation into a vortex state: in macroscopic
superfluids, the state for $L=N$ would have a condensate of unit angular
momentum and would be lower in energy than neighboring ground states. In the
weak interaction limit ($gN=0.096\ll 1$), our results only show slightly
deviation from that with LLL approximation. However, with increasing the
interaction strength, the interaction energy $V_{int,L}$ drops significantly
for each angular momentum. This implies a spatial reconfiguration of bosons
since they should push away from each other to reduce the strong interaction
as $g$ increases.

We next investigate the density distribution of bosons for various
interaction strengths as shown in figure 3. In general, the density is found
to become decreasingly small and broad as $g$ increases. For all angular
momenta, the shape of curves for the weak and relatively strong interaction
strength differs largely: not only the maximum density of the former case is
nearly two times larger than that of the latter one, but the maximum
position is also shifted strongly. Moreover, when the angular momentum is
closer to $L=N$, a maximum surrounded by an out ring of lower density
appears at the $r=0$ for a relatively strong interaction strength $g=0.3$
(see figure 3c).

To further examine the structure of the ground state, we calculate the
conditional probability distributions (CPDs), which measure the density
correlation among bosons \cite{liuxj98}. Unlike the usual density
distribution, which is {\em circularly symmetric} under the rotation
invariant confinement, the CPD reflects an {\em intrinsic} density
distribution of bosons \cite{cpd-note}. And thus it is indeed an observable
quantity in experiments instead of the circularly symmetric density
distribution. This is best illustrated for a condensed boson gas, where the
circularly symmetry is broken spontaneously, and CPD is then related to the
measurement of boson density \cite{cpd-note}. Note also that the CPD has
been extensively used in the study of electron correlations in doubly
excited helium-like atoms \cite{ezra} and analysis of the formation of
Wigner molecule in quantum dots \cite{yl}. We define the CPD for finding one
boson at ${\bf r}$ given that another is at ${\bf r}_0$ as 
\begin{equation}
{\cal P}({\bf r\mid r}_0)=\frac{\left\langle \Psi _{GS}\right|
\sum\limits_{i\neq j}\delta ({\bf r}-{\bf r}_i)\delta ({\bf r}_0-{\bf r}%
_j)\left| \Psi _{GS}\right\rangle }{(N-1)\left\langle \Psi _{GS}\right|
\sum\limits_j\delta ({\bf r}_0-{\bf r}_j)\left| \Psi _{GS}\right\rangle },
\end{equation}
where $\left| \Psi _{GS}\right\rangle $ represents the ground state. In
figure 4, we show a plot of CPDs for some selected angular momenta and for a
system of $N=16$ bosons. With increasing the two-body interaction strength,
the bosons spread more widely in x-y plane as expected, in line with density
distributions shown in figure 3. On the other hand, the evolution of
producing a single vortex state at $N=L$ does not change significantly even
as $g$ is varied over several ($\sim 2$) orders of magnitude.

In conclusion, we have studied the ground state of a repulsively interacting
Bose-Einstein condensate with a nonvanishing angular momentum. We found that
all physical quantities, especially the effective interaction energy, are
strongly affected by a dimensionless interaction strength $g$. We realize
that our result is based on a few-Boson system, and its validity to a
Bose-Einstein condensate should be further checked by more rigorous analytic
and numerical treatments, such as the quantum Monte-Carlo simulation. We
also note that although the so-far published experiments have mainly studied
systems that do not satisfy the condition $gN\sim 1$ or $gN\ll 1$, systems
that do satisfy this condition are accessible with current experimental
techniques \cite{feshbach}.

We would like to thank Y. Zhou and Y.-X. Miao for their stimulating
discussions. X. Liu was supported by NSF-China (Grant No. 19975027 and
19834060). H. Hu acknowledges the support of Profs. Jia-Lin Zhu and
Jia-Jiong Xiong, and a research grant from NSF-China (Grant No. 19974019).

\begin{center}
{\bf Figures Captions}
\end{center}

Fig.1. Ground state energy of a single vortex state versus the cutoff energy
for a system of $N=10$ bosons at the moderate (a) and relatively strong (b)
interaction strengths. The energy is measured in units of $\hbar \omega $.
Both the two cases show convergence at $E_c>N+L+10$.\newline

Fig.2. $V_{int,L}$ (in units of $\hbar \omega $) as a function of angular
momentum for different two-body interaction strength. In the LLL
approximation, it is described precisely by an algebraic expression $%
V_{int,L}=N(N-L/2-1)$ for $2\leqslant L\leqslant N$ \cite{bertsch,analytic}.%
\newline

Fig.3. Density distribution of $N=16$ bosons for various angular momenta.
The unit of length is the oscillator length $a=\sqrt{\hbar /M\omega }$. The
short dashed, dotted, dash dotted lines correspond to $g=0.006$, $0.1$, and $%
0.3$, respectively. The results with LLL approximation is also delineated by
a solid line for comparison. \newline

Fig.4. Selected conditional probability distributions for a system of $N=16$
bosons. The value of x, y in each subplot ranges from $-3.0a$ to $+3.0a$. $%
{\bf r}_0=(0,1.0a)$, and other choice of ${\bf r}_0$ does not alter our
results qualitatively.


\begin{references}
\bibitem{review}  F. Dalfovo, S. Giorgini, L. P. Pitaevskii, and S.
Stringari, Rev. Mod. Phys. {\bf 71}, 463 (1999).

\bibitem{matthews}  M. R. Matthews, B. P. Anderson, P. C. Haljan, D. S.
Hall, C. E. Wieman, and E. A. Cornell, Phys. Rev. Lett. {\bf 83}, 2498
(1999).

\bibitem{madison}  K. W. Madison, F. Chevy, W. Wohlleben, and J. Dalibard,
Phys. Rev. Lett. {\bf 84}, 806 (2000); F. Chevy, K. W. Madison, and J.
Dalibard, Phys. Rev. Lett. {\bf 85}, 2223 (2000); K. W. Madison, F. Chevy,
W. Wohlleben, and J. Dalibard, e-print cond-mat/0004037 (2000).

\bibitem{rokhsar}  D. Rokhsar, Phys. Rev. Lett. {\bf 79}, 2164 (1997).

\bibitem{sinha}  S. Sinha, Phys. Rev. A {\bf 55}, 4325 (1997).

\bibitem{feder}  D. L. Feder, C. W. Clark and B. I. Schneider, Phys. Rev.
Lett. {\bf 82}, 4956 (1999).

\bibitem{fetter}  A. A. Svidzinsky and A. L. Fetter, Phys. Rev. Lett. {\bf 84%
}, 5919 (2000).

\bibitem{marzlin}  K.-P. Marzlin, W. Zhang, and B. C. Sanders, Phys. Rev. A 
{\bf 62}, 013602 (2000).

\bibitem{wilkin98}  N. K. Wilkin, J. M. F. Gunn, and R. A. Smith, Phys. Rev.
Lett. {\bf 80}, 2265 (1998).

\bibitem{mft}  R. A. Butts and D. S. Rokhsar, Nature {\bf 397}, 327 (1999).
G. M. Kavoulakis, B. Mottelson, and C. J. Pethick, Phys. Rev. A {\bf 62},
063605 (2000).

\bibitem{mottelson}  B. Mottelson, Phys. Rev. Lett. {\bf 83}, 2695 (1999).

\bibitem{bertsch}  G. F. Bertsch and T. Papenbrock, Phys. Rev. Lett. {\bf 83}%
, 5412 (1999); T. Papenbrock and G. F. Bertsch, Phys. Rev. A {\bf 63},
023616 (2001).

\bibitem{wilkin00}  N. K. Wilkin and J. M. F. Gunn, Phys. Rev. Lett. {\bf 84}%
, 6 (2000). N. R. Cooper and N. K. Wilkin, Phys. Rev. B {\bf 60}, 16279
(1999).

\bibitem{mft-vs-ex}  A. D. Jackson, G. M. Kavoulakis, B. Mottelson, and S.
M. Reimann, Phys. Rev. Lett. {\bf 86}, 945 (2001).

\bibitem{analytic}  A. D. Jackson and G. M. Kavoulakis, Phys. Rev. Lett. 
{\bf 85}, 2854 (2000). R. A. Smith and N. K. Wilkin, Phys. Rev. A {\bf 62},
061602 (2000). T. Papenbrock and G. F. Bertsch, e-print cond-mat/0008286
(2000). Wen-Jui Huang, Phys. Rev. A 63, 015602 (2001).

\bibitem{ahsan}  M. A. H. Ahsan and N. Kumar, e-print cond-mat/0011212
(2000).

\bibitem{isoshima}  T. Isoshima and K. Machida, J. Phys. Soc. Jpn. {\bf 68},
487 (1999).

\bibitem{nm-note}  We have checked numerically that the precision of our
results is only slightly improved by including more single particle states.

\bibitem{davidson}  E. R. Davidson, J. Comput. Phys. {\bf 17}, 87 (1975). A.
Stathopoulos and C.F. Fischer, Comput. Phys. Commun. {\bf 79}, 268 (1994).

\bibitem{rb}  P. S. Julienne, F. H. Mies, E. Tiesinga, and C. J. Williams,
Phys. Rev. Lett. {\bf 78}, 1880 (1997).

\bibitem{liuxj98}  X. Liu, H. Hu, L. Chang, W. Zhang, S.-Q. Li, and Y.-Z.
Wang, e-print cond-mat/0012259, to appear in Phys. Rev. Lett. (2001).

\bibitem{cpd-note}  The difference between CPD and usual boson density can
be naturally interpreted as follows: the CPD describes bosons in their {\em %
intrinsic} ({\em body-fixed}) frame of the reference, while the density
distribution describes bosons in the laboratory frame of reference where the
rotational and center-of-mass displacements are superimposed upon the
intrinsic probability density. However, in the mean-field treatment (under
an assumption of spontaneous symmetry breaking), the many-body condensate
wavefunction is taken to be the product of the single-particle state $\Psi
_{GS}({\bf r}_1,{\bf r}_2,...,{\bf r}_N)=\prod\limits_{i=1}^N\psi ({\bf r}%
_i),$ and the corresponding CPD is simply reduced to the density
distribution of bosons ${\cal P}({\bf r\mid r}_0)=\left| \psi ({\bf r}%
)\right| ^2.$

\bibitem{ezra}  G. S. Ezra and R. S. Berry, Phys. Rev. A {\bf 28}, 1974
(1983); see also R. S. Berry in {\it Structure and Dynamics of Atoms and
Molecules: Conceptual Trends}, edited by J. L. Calais and E. S. Kryachko
(Kluwer, Dordrecht, 1995) p. 155, and references therein.

\bibitem{yl}  C. Yannouleas and U. Landman, Phys. Rev. Lett. {\bf 85}, 1726
(2000); Phys. Rev. B {\bf 61}, 15895 (2000).

\bibitem{feshbach}  S. Inouye, M. R. Andrews, J. Stenger, H.-J. Miesner, D.
M. Stamper-Kurn, and W. Ketterle, Nature {\bf 392}, 151 (1998).
\end{references}
\end{document}